\documentclass[iop]{emulateapj}

\usepackage{graphicx}

\shorttitle{Light bending scenario in X-ray polarimetry}
\shortauthors{Dov\v{c}iak, Muleri, Goosmann, Karas \& Matt}

\newcommand{\msun}{M_{\odot}}
\newcommand{\dd}{{\rm d}}
\newcommand{\lamp}{{\rm lamp}}
\newcommand{\prim}{{\rm prim}}
\newcommand{\oo}{{\rm o}}
\newcommand{\inc}{{\rm inc}}
\newcommand{\refl}{{\rm refl}}
\newcommand{\disc}{{\rm disk}}
\newcommand{\cont}{{\rm cont}}
\newcommand{\lin}{{\rm line}}
\begin{document}

\title{Light bending scenario for accreting black holes in X-ray polarimetry}

\author{M.~Dov\v{c}iak\altaffilmark{1,5}}
\author{F.~Muleri\altaffilmark{2}}
\author{R.~W.~Goosmann\altaffilmark{3}}
\author{V.~Karas\altaffilmark{1}}
%\and
\author{G.~Matt\altaffilmark{4}}
\altaffiltext{1}{Astronomical Institute, Academy of Sciences,
Bo\v{c}n\'{\i}~II, CZ-14131 Prague, Czech~Republic}
\altaffiltext{2}{Istituto di Astrofisica Spaziale e Fisica Cosmica,
Via del Fosso del Cavaliere 100, I-00133 Rome, Italy}
\altaffiltext{3}{Observatoire Astronomique de Strasbourg, 11 rue de
l'Universit\'e, F-67000 Strasbourg, France}
\altaffiltext{4}{Dipartimento di Fisica, Universit\`a degli Studi
``Roma Tre'', Via della Vasca Navale 84, I-00146 Rome, Italy}
\altaffiltext{5}{dovciak@astro.cas.cz}

\begin{abstract}
We discuss a model of an X-ray illuminating source above an accretion disk of a
rotating black hole. Within the so called lamp-post scheme we compute the
expected (observed) polarization properties of the radiation reaching an
observer. We explore the dependencies on model parameters, employing Monte Carlo
radiation transfer computations of the X-ray reflection on the accretion disk
and taking general relativity effects into account. In particular, we discuss
the role of the black hole spin, of the observer viewing angle, and of the
primary X-ray source distance from the black hole. We give several examples of
the resulting polarization degree for two types of exemplary objects -- active
galactic nuclei and Galactic black holes. In order to assess potential
observability of the polarization features, we assume the sensitivity of the
proposed New Hard X-ray Mission (NHXM).

We examine the energy range from several keV to $\sim50\,$keV, so the iron-line
complex and the Compton hump are included in our model spectra. We find the
resultant polarization degree to increase at the higher end of the studied
energy band, i.e.\ at $\gtrsim20$~keV. Thus, the best results for polarimetry of
reflection spectra should be achieved at the Compton hump energy region. We also
obtain higher polarization degree for large spin values of the black hole, small
heights of the primary source, and low inclination angles of the observer.
\end{abstract}


\section{Introduction}
In active galactic nuclei (AGN) and Galactic black holes (GBH), X-rays
are produced near the black hole horizon \citep{FraKR02}. The spectrum
of these objects typically consists of several constituents, namely, a
multi-color thermal component arising from an optically thick accretion
disk and a power-law component, thought to originate in an optically
thin medium, a.k.a.\ a `corona' (e.g.,
\citeauthor{Bland90} \citeyear{Bland90};
\citeauthor{K99} \citeyear{K99}). A plausible scheme
suggests that the latter component is produced in a hot Comptonizing
layer above the disk. Various spectral features originate by the mutual
interaction of the accretion disk medium and the corona. In particular, very
prominent are the reflection features of iron, which occur in the 6--7
keV range in the local frame of the emitting material
\citep{Rem06,Fabian08}.

The mutual proportions among these components in a particular object
depend on the physical (spectral) state of the system, as well as the
intervening medium along the line of sight. These proportions are also
influenced by strong gravity near the black hole, where light bending
effects and energy shifts can play a significant role. Disentangling the
observed spectrum into its individual components is not a
straightforward task, as it poses a degenerate problem and does not
generally allow us to constrain all parameters. Nevertheless, accurate
determination of the spectral components is an essential step in
measuring the black hole spin, which is one of the most important
challenges in the present-day astrophysics of accreting black holes
\citep{Fabian04}. It has been proposed that X-ray polarimetry could
provide additional constraints on the accretion disk models, limit the
range of acceptable parameter values, and help us reducing the
ambiguities that cannot be resolved by spectroscopy alone
\citep{dovc08,li09,schn09}.

In the inner regions of black hole accretion disks, emerging photons are
strongly influenced by strong gravity \citep{Kato98}. The framework of
General Relativity (GR) is needed to properly determine the resulting
spectrum \citep{Fabian00,Karas06,Matt06,Mill07}. Relativistic effects include
the energy shifts, both due to special relativity (Doppler effect) and
GR (gravitational redshift), as well as the light bending aberration
effects that are particularly prominent near the photon orbit: $r_{\rm
ph}=3\,GM/c^2=4.43\,M/\msun\,$km for a non-rotating black hole (the radius
of $r_{\rm ph}$ decreases with the  black hole spin increasing). The
relevance of GR effects is even more pertinent for the calculations of
the observed polarimetric quantities \citep{conn77,conn80,laor90,Matt93,agol00,
dovc04b,dovc08,li09,schn09,schn10}.
This is mainly due to sensitivity of the polarimetry parameters to the
geometrical arrangement of the source.

In \cite{dovc04b}, we studied the GR effects on the polarization
properties of the radiation reflected by an accretion disk near a
rotating (Kerr) black hole. We considered illumination of the disk
surface by a hot corona. The corona acts, via the inverse Compton
effect, on thermal disk photons and it produces as an outcome  the
above-mentioned power-law component.

In this context, a special scheme has been dubbed the `lamp-post model'
\citep{Matt91, Martocchia96,Henri97,Martocchia00,Miniutti04,dovc04b}. According
to this model, the illumination is caused by a  point-like source
located on the rotation axis of the black hole, which is also the
accretion disk axis. This kind of a source can be identified, e.g., with
the base of a jet, although the exact interpretation of the `lamp' does
not need to be specified at this stage of the phenomenological scenario.
Generalizations of the lamp-post model to the case of extended sources,
off-axis illumination, and non-planar disks have been also investigated
\citep{Dabrowski01,Nied08}. Different, but also highly relevant and interesting
geometry of an extended corona has been recently studied by \citet{schn10},
taking into account relativistic effects and even the self-irradiation
of an accretion disk by its own returning radiation.

The motivation for the present paper arises from the advent of X-ray
polarimetry in the field of AGN and GBH
\citep{Bellazzini2006,Bellazzini2010}.  The paper revisits the light
bending scenario of accreting black holes and it improves the
calculations of the flux and polarization of the reflected disk
emission, with the emphasis on the observability of GR effects and the
possibility to constrain the model parameters with future X-ray
polarimeters on-board satellites, such as the currently proposed  New
Hard X-ray imaging and polarimetric Mission (NHXM)
\citep{Tagliaferri2010,Tagliaferri2010b}, or the confirmed Gravity and
Extreme Magnetism SMEX (GEMS) mission \citep{Swank10}.

The paper is organized as follows. In section \ref{sec:model}  we
briefly introduce the lamp-post scheme for black-hole accretion disks.
In section \ref{sec:primary} we describe the primary
component which acts as a source of the X-ray irradiation.
Here we also provide the relevant equations and we calculate the illumination
of the disk.
In section \ref{sec:reflected} we describe details of the  radiation transfer
computation that we apply in order to derive the reflection component.
These considerations are then employed in section \ref{sec:infinity},
where we discuss a consistent procedure connecting the primary source
position on the rotation axis near the horizon with the signal
observed at arbitrary direction far from the black hole.
Results for the predicted polarization properties are then
described in section \ref{sec:results}, taking into account the continuum
together with the iron line features. Finally, in section
\ref{sec:observations} we describe the observational prospects with the
currently planned X-ray polarimetric missions. We demonstrate the
observability of GR  polarimetric features by modeling the expected
data from two proto-typical sources and assuming the sensitivity of the
NHXM polarimeter as an example. Section \ref{sec:conclusions} summarizes
the main results and concludes the paper.

\section{The lamp-post model for black-hole accretion disks}
\label{sec:model} The lamp-post model was originally proposed by
\cite{Matt91} and \cite{Petrucci97} with the aim
to provide a simple common scheme for the origin of the X-ray power-law
continuum and the relativistic spectral features seen in accreting black
hole sources. A natural motivation for this scenario arises from the
presumed co-existence of gaseous medium in different states
in the inner regions of the accretion disks, and hence
the action of different radiation processes that shape the resulting
spectra. This relates, in
particular, to the spectrum expected from a geometrically thin,
optically thick accretion disk, surrounded by a hot, diluted corona,
possibly with an outgoing jet in some objects.

The light-bending scenario was applied by \cite{Miniutti04} to explain
the time behavior of the primary and the reprocessed emission of the
proto-typical Seyfert~1 galaxy, MCG--6-30-15, where the spectral
features from the reprocessing vary less than the primary X-ray
continuum. These authors notice that, if the source of primary
illumination is located very close to the black hole, the observed
spectral properties may result as consequence of rearrangements of the
geometrical proportions of the object. This can be interpreted in terms
of a point-like source, representing, for example, shocks in the base of
a jet, or the location where intense flares arise via magnetic
reconnection processes. \cite{Rossi05} showed, using RXTE data for the
X-ray transient XTE J1650-500, that the light bending scheme may well
apply also to Galactic black hole systems.

\begin{figure}[tbh!]
\begin{center}
\includegraphics[width=8cm]{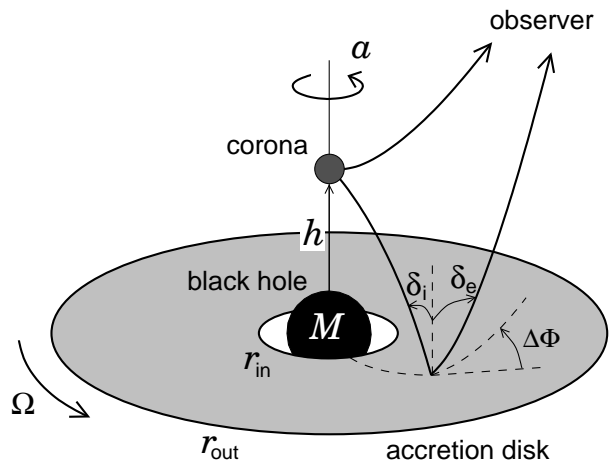}
\end{center}
\caption{Scheme of the model. An elevated primary source situated on
the axis of the black hole at height $h$ illuminates the corotating
accretion disk. The observer at infinity receives both the unpolarized
power-law radiation from the lamp-post as well as the polarized
reflected light from the disk. The incident angle, $\delta_{\rm i}$,
between the incoming light ray and the disk normal, the emission angle,
$\delta_{\rm e}$, between the outgoing light ray and the disk normal, and
the relative azimuthal angle, $\Delta \phi$, between the incident and
emitted light rays define the scattering geometry in the local frame
co-moving with the accretion disk. The normal to the disk and incident and
emitted light rays projected onto the disk surface are depicted by the dashed
lines.}
\label{fig:scheme}
\end{figure}

According to the original, most simple version of this model, the
primary `lamp' moves along the black hole axis, typically at heights of
only $\sim10$--$20$ gravitational radii above the disk plane. The
directly seen continuum component then changes its flux very
prominently, mainly due to the effect of changing the gravitational
redshift. On the other hand, the reprocessing spectral features are less
variable, therefore reproducing the observed pattern, which otherwise is
difficult to apprehend. This scheme is attractive because it allows us
to connect the black hole parameters to the observed spectral
variations, and so it has been pursued in a number of papers.

The above-mentioned interpretation, however, imposes the right
proportions between the source height and the accretion disk size, so
that the light-bending effect operates efficiently in the desired way.
This may not be generally satisfied, or it may be unlikely to
happen in majority of sources. The check can be performed by comparing
results of various approaches, such as timing, broad-band spectra, and
polarimetry. Recently, the {\tt rms} variability of MCG-6-30-15 was
investigated by \cite{nied10} and the Suzaku  broad-band spectra were
discussed by \cite{Zycki10}. These authors indeed find various
complications that appear when the simple lamp-post scenario is applied
to complex data.

The basic definition of the model set-up follows \cite{dovc04b}, see
Figure~\ref{fig:scheme}. A point-like emitting source on the black hole
axis is located at the height $h$. The primary emission that is assumed
to be unpolarized illuminates the accretion disk where it is reprocessed and
part of it is emitted towards the observer. In the lamp post scenario the
path of the light that travels from the lamp-post through some point in
the disk up to the observer is set for each position on the disk
unambiguously, thus defining the scattering geometry at every point in
the disk plane, see Figure~\ref{fig:scheme}. In this paper we consider only
direct photons, i.e., we assume optically thick disk and we neglect the
photons that might bend near the photon orbit in such a way that they
would circle around the black hole and only then either reach the
observer or strike the surface of the disk.

We introduce a number of improvements in addition to our previous
polarization computations \citep{dovc04b}:
\begin{itemize}
 \item The reflection emission is computed with a more sophisticated
       multi-scattering Monte Carlo code for the radiation transfer
       in the accretion disk. As a result of this modeling,
 \begin{itemize}
   \item the azimuthal dependence of the emission is taken into account
         (i.e., the reflection dependence on the azimuthal directions between
         incident and emitted light rays);
   \item the reflection is computed also for higher energies (up to 100~keV
         locally measured energy) and the resulting Compton hump is included in
         these new computations.
 \end{itemize}
 \item We have further refined the computation of light geodesics from the axis
       to the disk plane, and further to the observer, so additional
       improvements as opposed to our previous results were achieved, especially
 \begin{itemize}
   \item self-consistent computations for different spin values
         ($a=0.9987$ in our previous work);
   \item an extended range of possible heights of the
         primary source on the axis (we are able to trace
         the dependence on the height more precisely);
   \item geodesics from the axis to the observer are treated in a proper way,
         i.e., the lensing effect is included (however, we find that the
         previous approximation was already very good and this point does
         not lead to much different results).
 \end{itemize}
 \item We also include a contribution to the total signal from unpolarized
fluorescent K$\alpha$ and K$\beta$ lines.
\end{itemize}

We calculated the flux and polarization of the reprocessed emission, as
well as the flux of the direct emission, in four energy bands (2--6 keV,
6--10 keV, 10--20 keV, 20--50 keV) for three different inclination
angles (30, 60, and 80 degrees, respectively), and for various values of
the height $h$. Two different values of the angular momentum of the
black hole have been explored: a static, $a$=0, and an extremely
rotating black hole, $a$=1, where, as customary, $a$ indicates the
angular momentum per unit mass in units of the gravitational radius. We
use these units for the remainder of this paper. The accretion disk in our
computations extends from the marginally stable orbit ($r_{\rm in}=r_{\rm ms}$)
up to the radius $r_{\rm out}=1000\,GM/c^2$.

\section{Primary source, observed primary radiation, and illumination of
accretion disk}
\label{sec:primary}
We assume the primary source of unpolarized radiation to be a
point-like patch of the corona located at height $h<100$ above the black
hole (measured from the center). It radiates isotropically with the
specific intensity being the usual power-law dependence on the energy,
$I_\lamp(E)=N_\lamp\,E^{-\alpha}$. In our computations we use the value
of the power-law index $\alpha=1$ and the normalization factor $N_\lamp$
to be unity in the local static frame.

\begin{figure*}[tbh!]
\begin{center}
\includegraphics[width=0.95\textwidth]{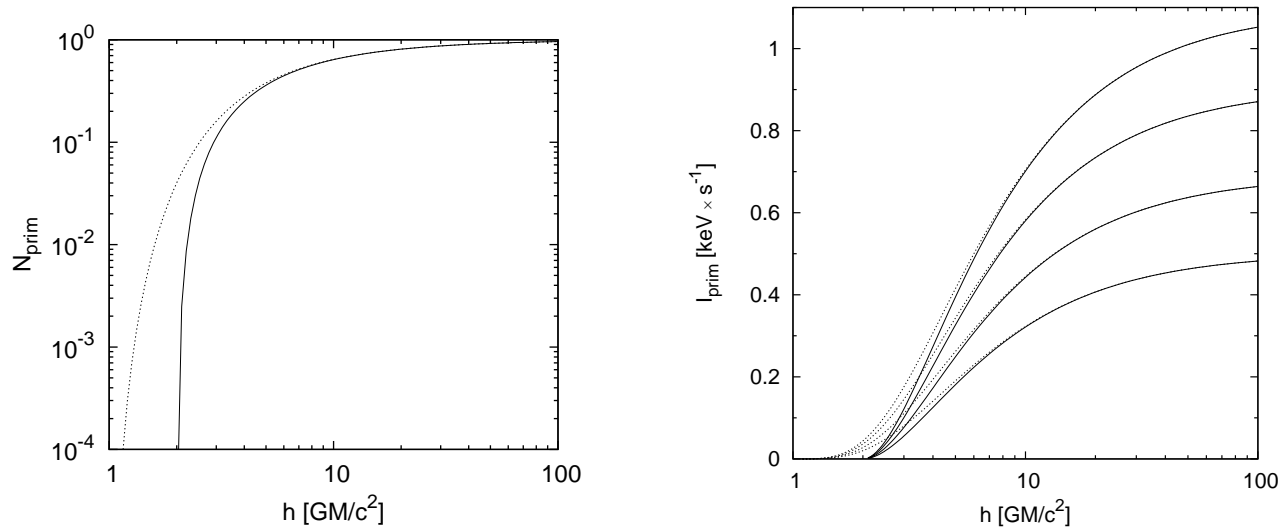}
\end{center}
\caption{Primary emission. {\em Left:} The emission received by the observer
from the primary source is diminished by the factor $N_\prim$ because of the
strong gravity of the central black hole. The dependence on the height of the
illuminating corona, $h$, is shown, if the intensity of the primary
power-law emission is $E^{-1}$. This factor is almost independent on the
inclination of the observer (the curves for inclinations $1^\circ$ and
$85^\circ$ would be separated by less than the width of the shown
curves). {\em Right:} The primary component of the intensity far from the source
(at radial infinity),
integrated in the energy ranges 2--6, 6--10, 10--20, and 20--50~keV
(curves with higher to lower normalization). The dependence on the
height of the primary source is shown with the solid curves for the
non-rotating Schwarzschild black hole (a=0) and with the dotted curves
for the maximally rotating Kerr black hole (a=1).}
\label{fig:I_prim}
\end{figure*}

The primary emission received by the observer gets diluted by the
relativistic effects that act on photons on their path, mainly near the
central compact body. The observed intensity, $I_\prim(E)\equiv
\dd E/(\dd t\,\dd\Omega\,\dd\nu)$, will be again a power-law in
the observed energy
\begin{equation}
\label{eq:I_prim}
I_\prim(E)=N_\prim\,E^{-\alpha}\ ,
\end{equation}
where
\begin{eqnarray}
\nonumber
N_\prim \ & = & \ g_\prim^{\alpha+1}\,\frac{\dd\Omega_\lamp}{\dd\Omega_\oo}\ =\\
 & = & \left(1-\frac{2h}{h^2+a^2}\right)^{\frac{\alpha+1}{2}}\,
\frac{\sin{\theta_\lamp}}{\sin{\theta_\oo}}\,
\frac{\dd\theta_\lamp}{\dd\theta_\oo}\ ,
\end{eqnarray}
with $\theta_\lamp$ and $\theta_\oo$ being the local emission angle and
the inclination angle of the observer, respectively, both measured with
respect to the system axis. We denoted the energy shift (from $E_\lamp$
to $E_\prim$) of the primary photons detected by the observer by
$g_\prim=E_\prim/E_\lamp=1/U_\lamp^t(h,a)=-g_{tt}^{1/2}(h,a)$. Here,
$U_\lamp^\mu$ is the four-velocity of the motionless lamp-post and
$g_{\mu\nu}$ is the space-time metric. In Figure~\ref{fig:I_prim} we show
the dependence of the factor $N_\prim$ and intensity $I_\prim$
integrated in several energy bands on the height of the lamp-post for
Schwarzschild and extreme Kerr black holes.
One can see that the relativistic effects for direct radiation become
small for heights $h\gtrsim 100$ and the difference between extremely
rotating and non-rotating black holes are well visible only for heights
$h\lesssim 6$.

\begin{figure*}[tbh!]
\begin{center}
\includegraphics[width=\textwidth]{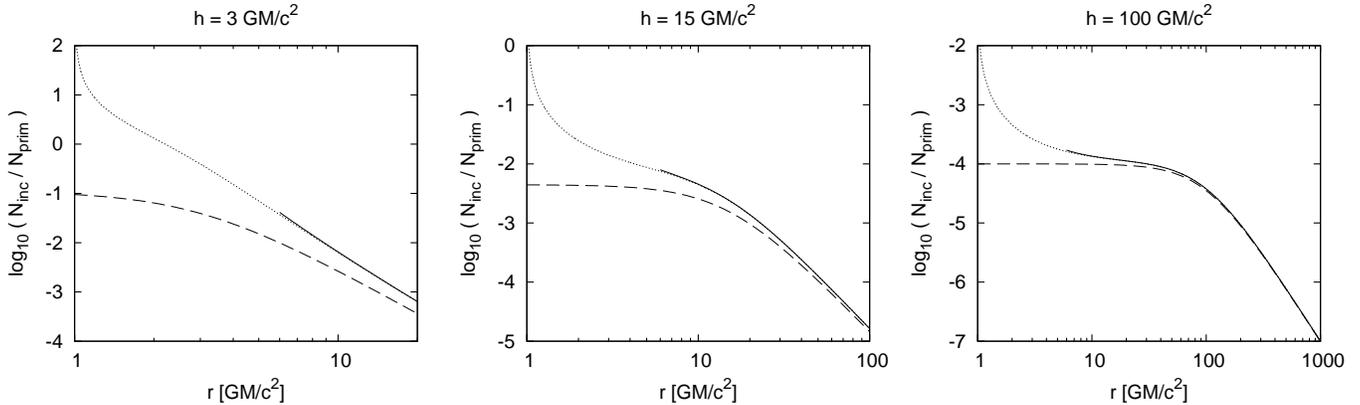}
\end{center}
\caption{Illumination of the disk. The radiation incident on the accretion disk
is affected by the
relativistic effects. The ratio of the incident radiation to the observed
primary emission is shown for three heights of the primary source, $h=3,\,15$,
and $100\,GM/c^2$ (from left to right). The dependence on the radius
is shown for the non-rotating Schwarzschild black hole (a=0) with the
solid curve, for the maximally rotating Kerr black hole (a=1) with the
dotted curve, and for the Newtonian case with the dashed curve.}
\label{fig:N_inc}
\end{figure*}

Due to the lamp-post geometry and relativistic effects, the primary
source illuminates the accretion disk unevenly. The reflected flux from
the disk is proportional to the incident flux on the disk. The incident
flux is also the power-law with the same index $\alpha$ and with the
following normalization factor \citep[see][]{dovc04b}\footnote{Note,
that we use the power-law index $\alpha$ for the specific intensity,
here, whereas in \cite{dovc04b} power-law index $\Gamma$ for the photon
number density flux was used.}
\begin{eqnarray}
\nonumber
N_\inc \ & = & \ \frac{g_\inc^\alpha}{U^t_\lamp}\,\frac{\dd\Omega_\lamp}{\dd S}
\ =\\
\nonumber
& = & \left(\frac{r^2+ar^{1/2}}{r\sqrt{r^2-3r+2ar^{1/2}}}\right)^{\alpha}\,
\left(1-\frac{2h}{h^2+a^2}\right)^{\frac{\alpha+1}{2}}\\[1mm]
\label{eq:N_inc}
& & \times\ \frac{\sin{\theta_\lamp}}{r}\,\frac{\dd\theta_\lamp}{\dd r}\ .
\end{eqnarray}
Here, the energy shift $g_\inc$ (from $E_\lamp$ to $E_\inc$) of the
primary photons incident on the accretion disk (measured in local frames) is
$g_\inc=E_\inc/E_\lamp=U^t_\disc/U^t_\lamp$ with $U^\mu_\disc$ being the
Keplerian four-velocity of the disk. The radial distance $r$ is the
Boyer-Lindquist radial coordinate (we use it here because eventually we
integrate in Boyer-Lindquist coordinates on the disk and not in the
local frame of the disk). We show the ratio of the incident radiation to
that received by the observer directly from the lamp-post in
Figure~\ref{fig:N_inc}.

Compared to the Newtonian case (see the dashed plot in the same
figure) this ratio is in both relativistic cases for spinning and
non-spinning black holes much larger in the inner accretion disk. The
difference between the relativistic and non-relativistic cases increases
with the decreasing height of the primary source of radiation. However,
we must recall that the incident radiation shown in the figures will
still, after it is reprocessed by the disk, undergo some changes on its way from
the disk to the observer caused
by the relativistic effects. The emission from the vicinity of the black
hole horizon will eventually be diminished by the redshift. As a result, for the
low lamp-post heights, the difference between the relativistic and
Newtonian ratio of the received direct emission to the received
reflected radiation will not be so large.

\section{Reflected emission}
\label{sec:reflected}
\begin{figure*}[tbh!]
\begin{center}
\includegraphics[width=\textwidth]{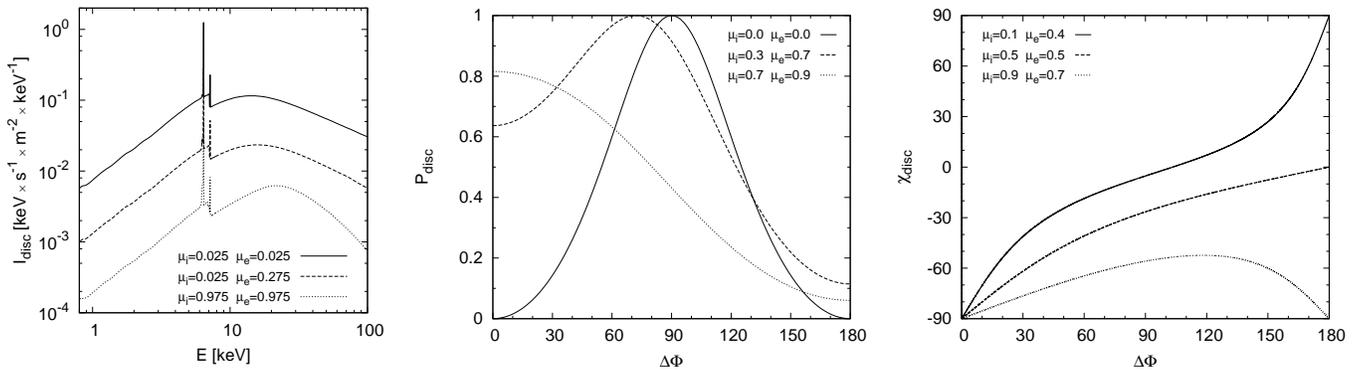}
\end{center}
\caption{Reprocessed radiation.
The properties of the local reprocessed radiation depend on the
position on the disk. This is due to their dependence on the local
scattering geometry. {\em Left:} The local intensity varies by more than
one order of magnitude for different values of incident and emission
angles. The shape of the Compton hump changes as well. Examples of
spectra for three different pairs of values of the incident and emission
cosines, $\mu_{\rm i}$ and $\mu_{\rm e}$, are shown. Here we assumed the
relative azimuthal angle between incident and emitted light rays to be
small ($\Delta\Phi=2.5^{\circ}$). {\em Middle:} The local polarization
degree can have any value between 0 and 1. We show the dependence on the
relative azimuthal angle $\Delta\Phi$ for three pairs of incident and
emission cosines. {\em Right:} The local polarization angle obtains any
value between $-90^\circ$ and $90^\circ$. The dependence on the relative
azimuthal angle $\Delta\Phi$ for three pairs of incident and emission
cosines is shown. A polarization angle of $0^\circ$ represents the direction
perpendicular to the disk.}
\label{fig:local_radiation}
\end{figure*}

In the equatorial plane of the system we assume a neutral cold
geometrically thin and optically thick Keplerian accretion disk. Thus, neutral
Fe fluorescent K$\alpha$ and K$\beta$ lines are
present in the local spectra together with the iron edge.
Intensity of the reflected radiation in the local frame co-moving
with the accretion disk was computed by the Monte Carlo multi-scattering code
{\tt NOAR} \citep{Dumont00}. Spectral features caused by scattering are hence
automatically included in the model spectra, namely, the Compton hump is
present; it occurs at energies
typically $\simeq15$--50~keV (see example spectra in the left panel of
Figure~\ref{fig:local_radiation}).

The normalization of the local flux as
well as its shape (mainly in the Compton hump region) depend on the
local scattering geometry, i.e., the incident and the emission angles, as
well as the relative azimuthal angle of the incident and emitted
light rays. We stress the dependence on the incident and
relative azimuthal angles of the lamp-post set-up. This feature is
additional to the pure directionality dependence on the emission angle in the
case of an extended corona when the disk is usually assumed to be illuminated
isotropically.

The single scattering approximation \citep{chan60} is used for the local
polarization of the reflected continuum component
\citep[for details see][]{dovc04b}. The line flux and the primary radiation are
supposed to be unpolarized. The polarization properties of the reflected
radiation depend highly on the scattering geometry, as is illustrated
in Figure~\ref{fig:local_radiation}.

The geometry of scattering is determined by the position where reprocessing
happens on the disk. In this set-up the incident and emission angles are decided
by the incoming and outgoing geodesics. Throughout the disk all three
angles, $\delta_{\rm i}, \delta_{\rm e}$, and $\Delta \phi$
(Figure~\ref{fig:scheme}), may attain almost all possible values. Thus
also the local polarization degree $P_\disc$ and angle $\chi_\disc$ may
achieve the whole range of values, 0 to 1 and $-90^\circ$ to $90^\circ$,
respectively, as is illustrated in Figure~\ref{fig:local_radiation}.
In our computations, a polarization angle of $0^\circ$ always represents the
direction perpendicular to the disk and it is rotated counter-clockwise for
positive values when looking towards the approaching photon.

\section{Radiation far from source}
\label{sec:infinity}
To get the resultant polarization far from the source, i.e.\ at radial infinity,
one has to integrate the
emission over the disk surface and thus one integrates the local
polarization properties. From the previous section it is obvious that
the result of such integration is not easily estimated because of the
complicated dependence on the local geometry of scattering. Moreover,
relativistic effects enhance radiation from some parts of the disk.
They also rotate the polarization angle, and thus the overall
observed polarization properties will be given by an interplay of the local
polarization properties on the disk and relativistic change acting on
photons on their way to the observer.

From our previous studies of polarization of thermal radiation of the
accretion disk \citep[see][]{dovc08} one expects relativistic depolarization of
the local polarization. This depolarization should be larger for a
spinning black hole when the inner part of the accretion disk extends
closer to the horizon, and thus is larger in size when compared to the
Schwarzschild non-rotating case where there is a large central hole in
the disk. It is in this inner part where the relativistic rotation of
the polarization angle is the most variable. Yet, in the lamp-post
geometry we may get different results (i.e. higher polarization for
higher spin) when the primary radiation is taken into account as well,
as we will see later on.

\begin{figure*}[tbh!]
\begin{center}
\includegraphics[width=\textwidth]{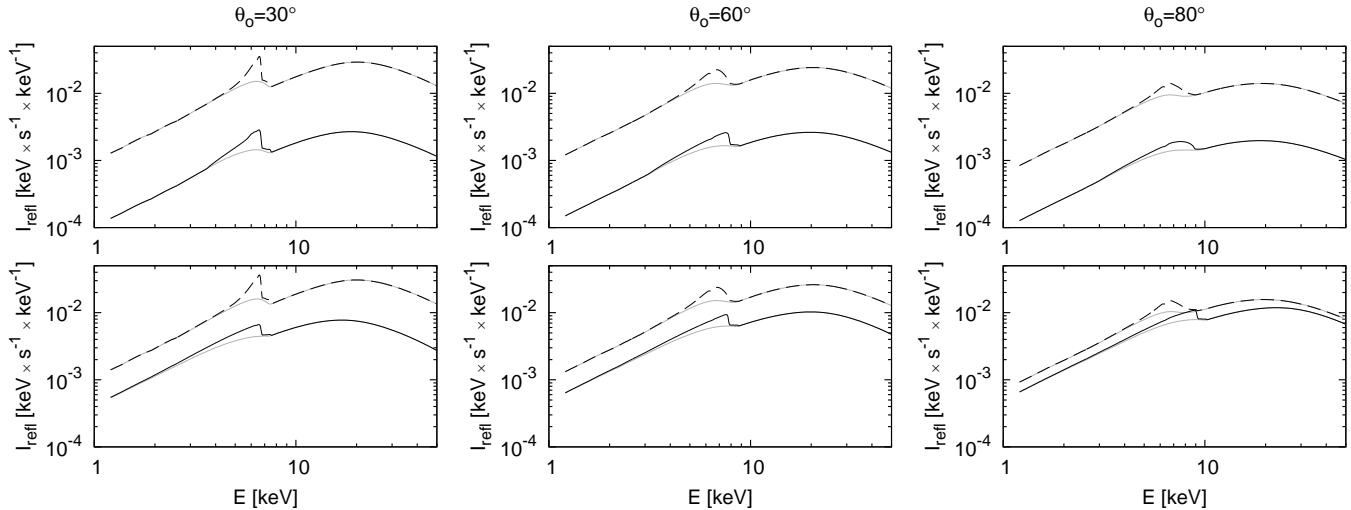}
\end{center}
\caption{Reflected component at infinity. The spectrum of the reflected
component as the observer at
infinity would measure it for three different inclination angles,
$\theta_{\rm o}=30^\circ,\,60^\circ$, and $80^\circ$ (from left to right)
and two heights of the primary source, $h=3\,GM/c^2$ (solid) and
$h=15\,GM/c^2$ (dashed). The top panels correspond to the
Schwarzschild case (a=0), the bottom panels correspond to the extremely
rotating Kerr case (a=1). The grey lines depict the spectrum when the Fe
K$\alpha$ and K$\beta$ lines are omitted.}
\label{fig:intensity_ref}
\end{figure*}

\begin{figure*}[tbh!]
\begin{center}
\includegraphics[width=\textwidth]{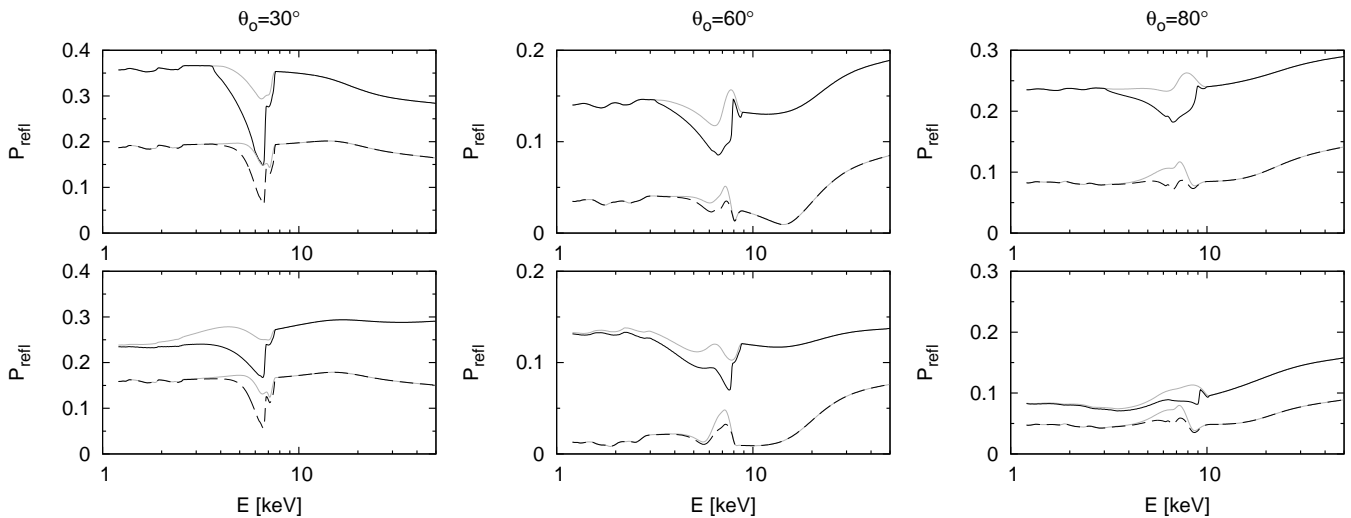}
\end{center}
\caption{Polarization of the reflected component.
The polarization of the reflected component as the observer at
infinity would measure it for three different inclination angles,
$\theta_{\rm o}=30^\circ,\,60^\circ$, and $80^\circ$ (from left to right)
and two heights of the primary source, $h=3\,GM/c^2$ (solid) and
$h=15\,GM/c^2$ (dashed). The energy dependence of the polarization
degree is shown. The top panels correspond to the Schwarzschild case
(a=0), the bottom panels correspond to the extremely rotating Kerr case
(a=1). The grey lines depict the polarization degree if the unpolarized
Fe K$\alpha$ and K$\beta$ lines are not considered.}
\label{fig:poldeg_ref}
\end{figure*}

The overall polarization at infinity is given by the Stokes parameters
that have to be integrated over the accretion disk surface,
\begin{eqnarray}
\nonumber
I_\oo \ & = & \ I_\refl+I_\prim =\\
\label{stokesI}
& = & \int_\Sigma \dd S\ G\,(I_\disc^{\cont}+I_\disc^{\lin})+I_\prim\ ,\\[2mm]
\label{stokesQ}
Q_\oo & = & \int_\Sigma \dd S\ G\,P_\disc\ I_{\disc}^{\cont}\,
                \cos{2(\chi_{\disc}+\psi)}\ ,\\[2mm]
\label{stokesU}
U_\oo & = & \int_\Sigma \dd S\ G\,P_\disc\ I_{\disc}^{\cont}\,
                \sin{2(\chi_{\disc}+\psi)}\ .
\end{eqnarray}
Here, the relativistic effects are expressed in terms of the transfer
function $G$ and the relativistic rotation of the polarization angle
$\psi$. For their definition and behavior see our previous work in
\cite{dovc04,dovc04a}. We have divided the intensity measured by the
observer into the part coming directly from the primary source,
$I_\prim$, and the part reflected by the accretion disk, $I_\refl$. The latter
one can still be divided into the polarized continuum part, $I_\refl^\cont$,
and the unpolarized intensity coming from fluorescent neutral iron lines
(K$\alpha$ and K$\beta$), $I_\refl^\lin$. We denote their counterparts
in the local frames in which the radiation is emitted, i.e. before
applying relativistic changes, as $I_\lamp$, $I_\disc^\cont$, and
$I_\disc^\lin$, respectively. The observed linear polarization degree
$P_\oo$ and polarization angle $\chi_\oo$ are computed from the Stokes
parameters in the usual way, i.e. $P_\oo=\sqrt{Q_\oo^2+U_\oo^2}/I_\oo$
and $\tan{2\chi_\oo}=U_\oo/Q_\oo$.

\begin{figure*}[tbh!]
\begin{center}
\includegraphics[width=\textwidth]{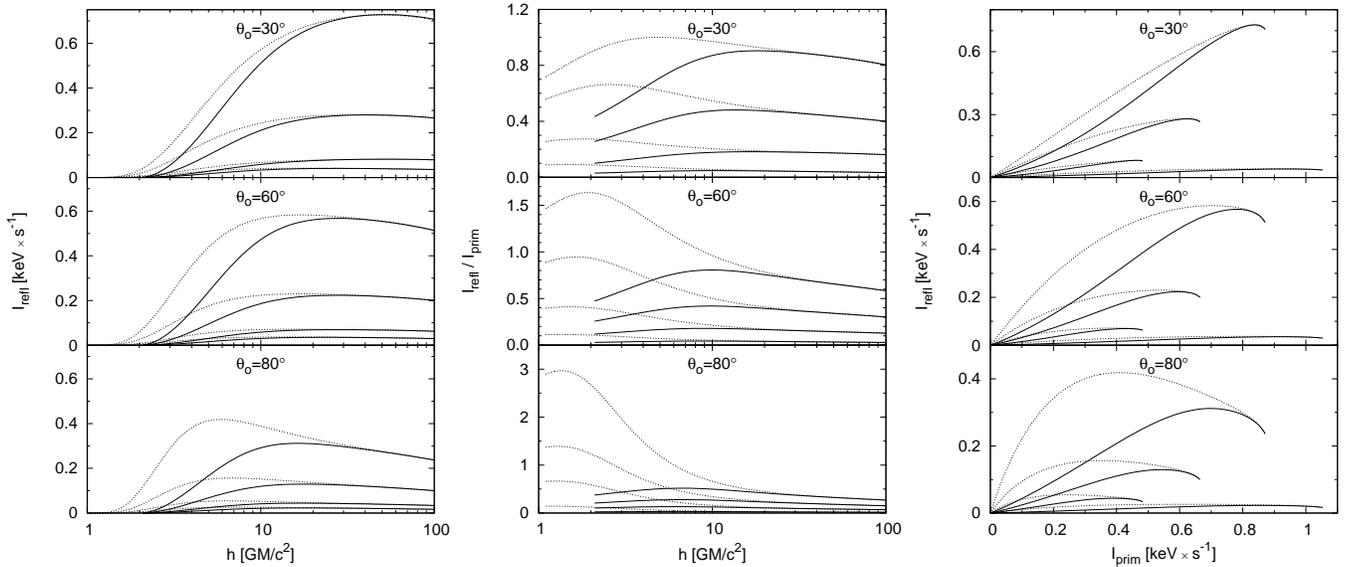}
\end{center}
\caption{Reflected intensity at infinity.
{\em Left:} The reflected component of the intensity
at infinity integrated in the energy ranges 2--6, 6--10, 10--20, and
20--50~keV (curves with lower to higher normalization) for three
different observer inclinations, $\theta_{\rm o}=30^\circ$, $60^\circ$
and $80^\circ$ (top to bottom panels). The dependence on the height of
the primary source is shown by the solid curve for the non-rotating
Schwarzschild black hole (a=0) and by the dotted curve for the maximally
rotating Kerr black hole (a=1). {\em Middle:} The same as on the left
but for the ratio of the reflected to the primary components of the
intensity at infinity. {\em Right:} The same as on the left but for the
dependence on the intensity of the primary source as an observer would
measure it at infinity.In this graph, the height of the primary source
increases with the primary intensity from $0$ to $100\,GM/c^2$.}
\label{fig:intensity_ref_integrated}
\end{figure*}

\section{Results}
\label{sec:results}
\subsection{Polarization of the reprocessed radiation and iron line features}

First, we will look at the polarization properties of the
relativistically broadened iron line. To this purpose we show only the
reflected component of the intensity $I_\refl$ and polarization degree
$P_\refl$ (Figures~\ref{fig:intensity_ref} and \ref{fig:poldeg_ref}). The
value of the polarization angle does neither depend on the primary
radiation nor on the intensity of the iron lines, thus we will consider
it only in the next section when we will discuss polarization properties
with the primary emission taken into account as well.

In Figure~\ref{fig:intensity_ref} one can see typical relativistic
reflection spectra with a broad iron line below $10\,$keV and Compton
hump that peaks around $20\,$keV. Here we show several examples from the
lamp-post geometry varying different model parameters -- the height of
the primary source, the inclination of the observer, and the spin of the
black hole. We show also the reflection continuum so that the intensity
of the line is emphasized. The line is broader for lower $h$ of the
lamp-post because more photons hit the disk near its inner edge (see
Figure~\ref{fig:N_inc}) where the line is shifted to the lower energy.

Due to the fact that the K$\alpha$ and K$\beta$ lines are unpolarized in
our model, the polarization degree $P_\refl$ should be reduced at the
energies to which the line is shifted. However, in some cases this needs
not to be true and there even can be a rise in polarization degree, see
the graphs for heights $h=15\,GM/c^2$ in Kerr case in
Figure~\ref{fig:poldeg_ref}. The reason for this behavior is the presence
of the iron edge, where the polarized continuum intensity suddenly
drops. As this edge is shifted to different energies there is a change
in the contribution of the polarized flux at the energies near it. The
overall effect can be either enhancing the polarization or decreasing it
(see the grey graphs representing only the continuum emission).

The polarization of the reflected component can be quite high. In our
example figures it is the highest at lower energies, $E\lesssim 4\,$keV,
for lower heights of the primary source, $h=3\,GM/c^2$, low
inclination, $\theta_\oo=30^\circ$, and for the Schwarzschild black hole
when it is slightly above $35$\%.

Next we show the reflected emission
(Figure~\ref{fig:intensity_ref_integrated}) as a function of the lamp-post
height, for the different energy bands (2--6, 6--10, 10--20, and
20--50~keV), and for the three values of the inclination angle
($30^\circ$, $60^\circ$, and $80^\circ$). The reflected radiation first
increases with the height, mainly due to the increase of the disk solid
angle as seen by the source (i.e. the light-bending effect decreases).
Thus for small heights the incident radiation on the disk increases with
the height. The decrease at large heights (above $h\gtrsim
30\,GM/c^2$) is primarily due to the finite size of the disk in our
computations (we used $r_{\rm out}=1000\,GM/c^2$). However, even if
the disk had extended into infinity, there would still have been slight
decrease in the reflected intensity with the height, as opposed to the
Newtonian case when it should stay constant. This is due to the fact
that the reflected radiation from the inner disk is more intense in the
relativistic case and as the lamp-post height increases the solid angle
of the inner region, where relativity effects are important, decreases.

\begin{figure*}[tbh!]
\begin{center}
\includegraphics[width=\textwidth]{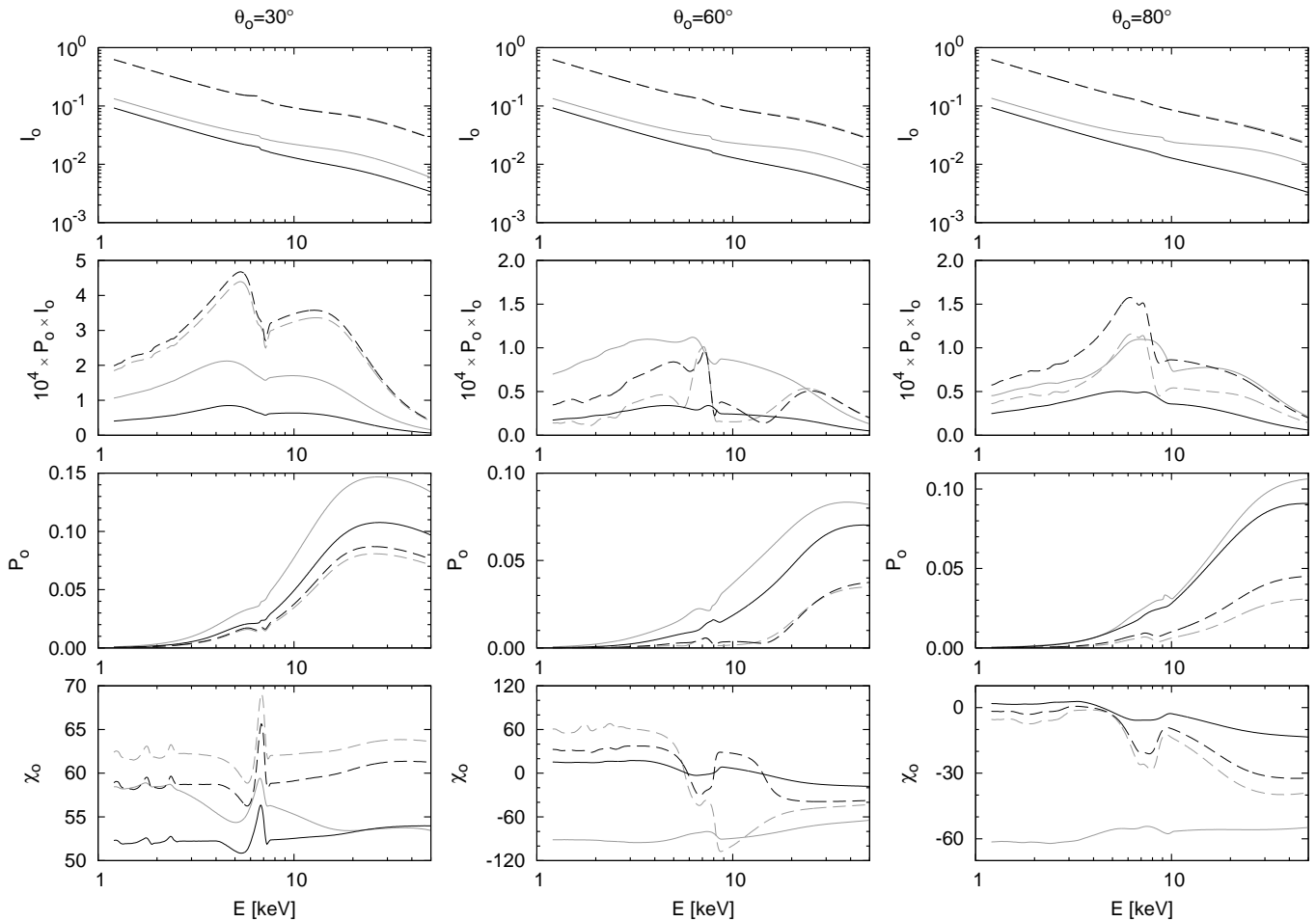}
\end{center}
\caption{Properties of the radiation at infinity.
Intensity, polarized intensity, polarization degree, and
polarization angle (top to bottom) at infinity for the total radiation,
i.e. both primary and reflected components are included. The observer
inclination is $\theta_{\rm o}=30^\circ,\,60^\circ$, and $80^\circ$ (from
left to right) and the height of the primary source is $h=3\,GM/c^2$
(solid) and $h=15\,GM/c^2$ (dashed). The black graphs correspond to
the Schwarzschild case (a=0), the grey graphs correspond to the
extremely rotating Kerr case (a=1).}
\label{fig:polar_ener}
\end{figure*}

For very large heights the reflected component approaches the Newtonian
case. From Figure~\ref{fig:intensity_ref_integrated} (left and middle panels)
one can see that the
intensity for different spins is almost the same already for heights around
$30\,GM/c^2$, which is also due to the fact that at these heights the
innermost accretion disk already makes quite a small contribution to the
overall reflected radiation. From the figure it is also apparent that
the contribution of the inner disk to the total intensity for the
extreme Kerr black hole is the largest for high inclination angles. In
Figure~\ref{fig:intensity_ref_integrated} we also show the dependence of
the ratio of the reflected to the primary intensity on the lamp-post
height and reflected versus primary intensity. Note, that for low
heights the contribution of the reflected component may be even larger
than that of the primary one (in higher energy bands). All these effects
are qualitatively similar to those described in
\cite{Miniutti04},
who adopted a somewhat different geometry for the primary source (i.e.\
a ring centered on the black hole axis).

\subsection{Overall polarization far from source}

If we consider the primary radiation as well, the decrease of the
polarization degree in the iron line region is much less visible,
because the primary flux at these energies is still large. On the other
hand, at high energies, where the contribution from direct radiation is
relatively lower and where the Compton scattering increases the
contribution of the reflected radiation, the polarization degree
increases. We illustrate this behavior in Figure~\ref{fig:polar_ener}
for two values of the lamp-post heights ($h=3$ and $15\,GM/c^2$). We
also show the energy dependence of the polarization angle in the same
figure. As previously mentioned for the polarization degree of the
reflected component, one can see quite a large effect of the iron edge.

\begin{figure*}[tbh!]
\begin{center}
\includegraphics[width=\textwidth]{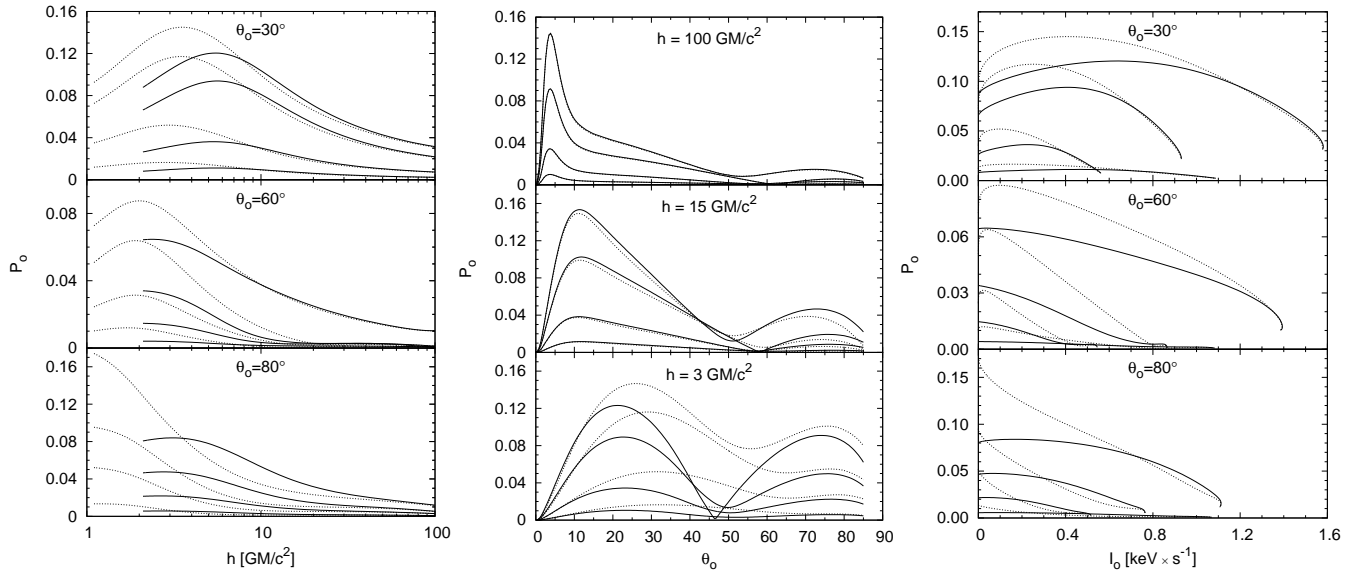}
\end{center}
\caption{Polarization degree at infinity. The values for the energy ranges
$2-6,\,6-10,\,10-20$, and $20-50\,$keV (curves with lower to higher
normalization) are shown. The solid curves are for the non-rotating
Schwarzschild
black hole (a=0), the dotted curves are for the maximally rotating Kerr
black hole (a=1). {\em Left:} The dependence on the height of the
primary source for three different observer inclinations, $\theta_{\rm
o}=30^\circ,\,60^\circ$, and $80^\circ$ (top to bottom panels). {\em
Middle:} The dependence on the observer inclination for three different
heights of the primary source, $h=100,\,15$, and $3\,GM/c^2$ (top to
bottom panels). {\em Right:} The dependence on the total intensity at
infinity for three different observer inclinations, $\theta_{\rm
o}=30^\circ,\,60^\circ$, and $80^\circ$ (top to bottom panels).
In this graph,
the height of the primary source increases with the total intensity from $0$ to
$100\,GM/c^2$.}
\label{fig:pol_degree_hie}
\end{figure*}

\begin{figure*}[tbh!]
\begin{center}
\includegraphics[width=\textwidth]{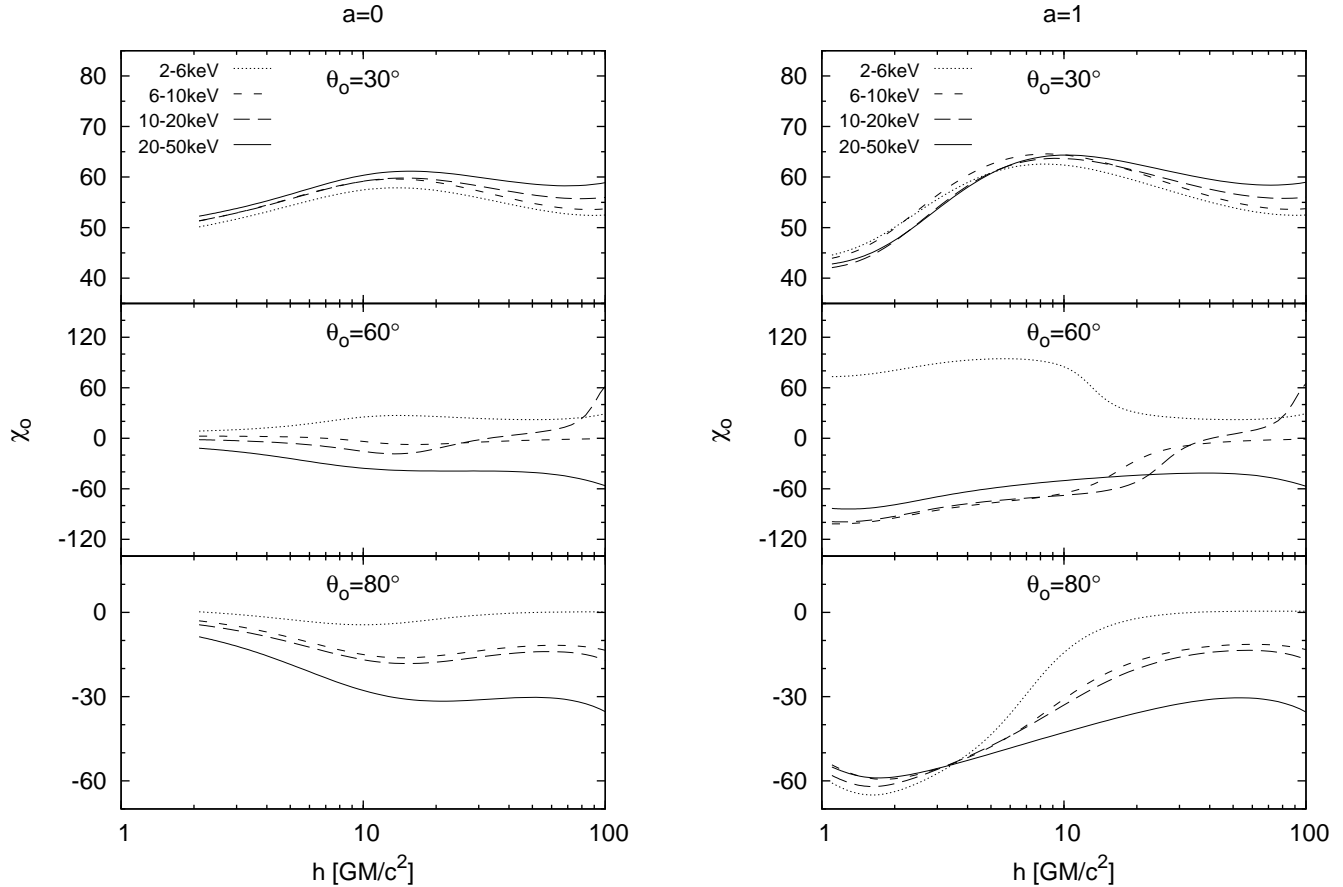}
\end{center}
\caption{Dependence of the polarization angle at infinity on the height of the
primary source. A polarization angle of $0^\circ$ represents the direction
perpendicular to the disk.
The values for the energy ranges
$2-6,\,6-10,\,10-20$, and $20-50\,$keV for three different observer
inclinations, $\theta_{\rm o}=30^\circ,\,60^\circ$, and $80^\circ$ (top
to bottom panels) are shown.
{\em Left:} The non-rotating Schwarzschild black hole case (a=0).
{\em Right:} The maximally rotating Kerr black hole case (a=1).}
\label{fig:pol_angle_he}
\end{figure*}

\begin{figure*}[tbh!]
\begin{center}
\includegraphics[width=\textwidth]{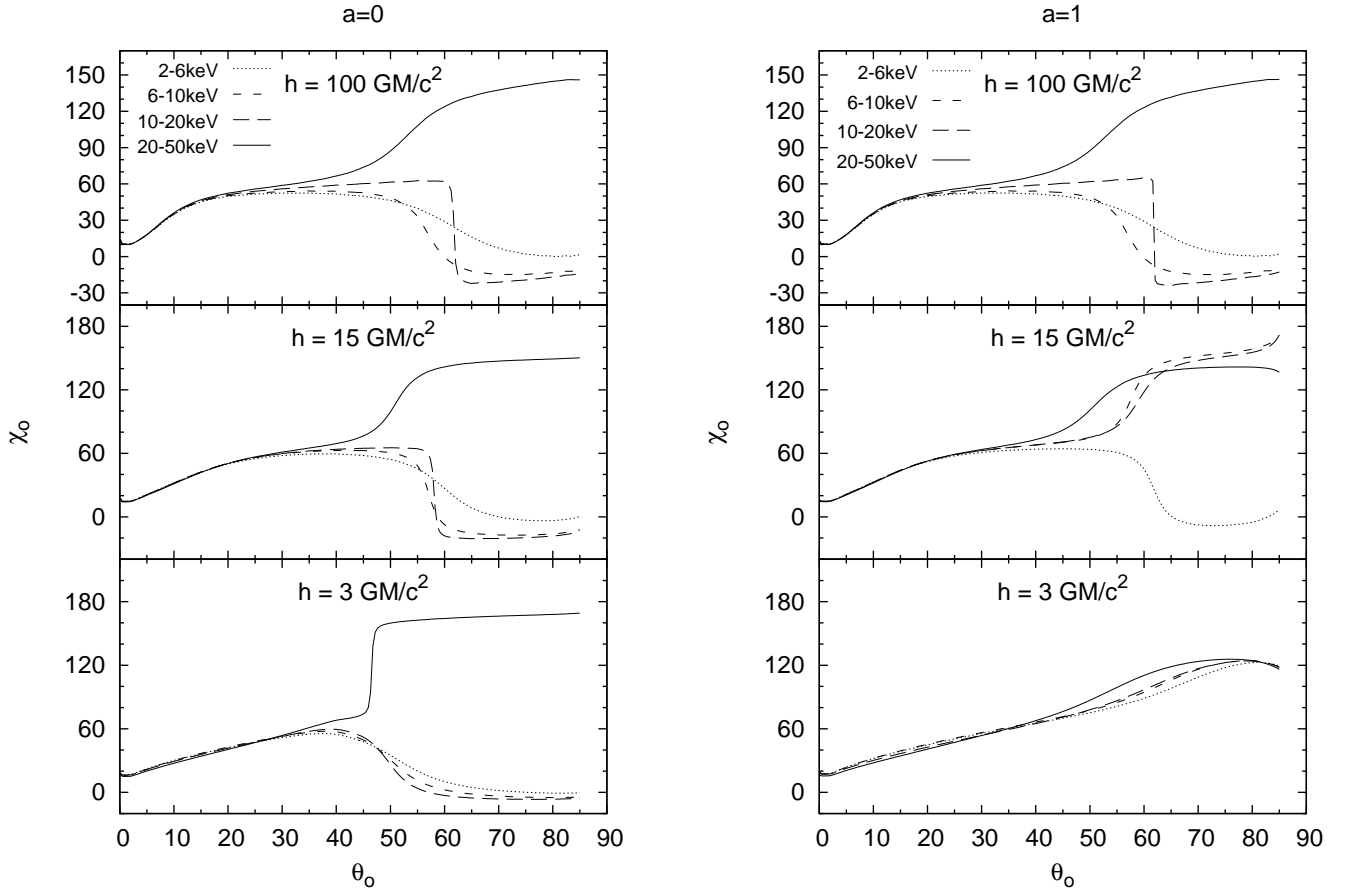}
\end{center}
\caption{Dependence of the polarization angle at infinity on the observer
inclination. A polarization angle of $0^\circ$ represents the direction
perpendicular to the disk.
The values for the energy ranges 2--6, 6--10, 10--20, and 20--50~keV for three
different heights of the primary source, $h=100$, $15$, and $3\,GM/c^2$
(top to bottom panels) are shown.
{\em Left:} The non-rotating Schwarzschild black hole case (a=0).
{\em Right:} The maximally rotating Kerr black hole case (a=1).}
\label{fig:pol_angle_ie}
\end{figure*}

\begin{figure*}[tbh!]
\begin{center}
\includegraphics[width=\textwidth]{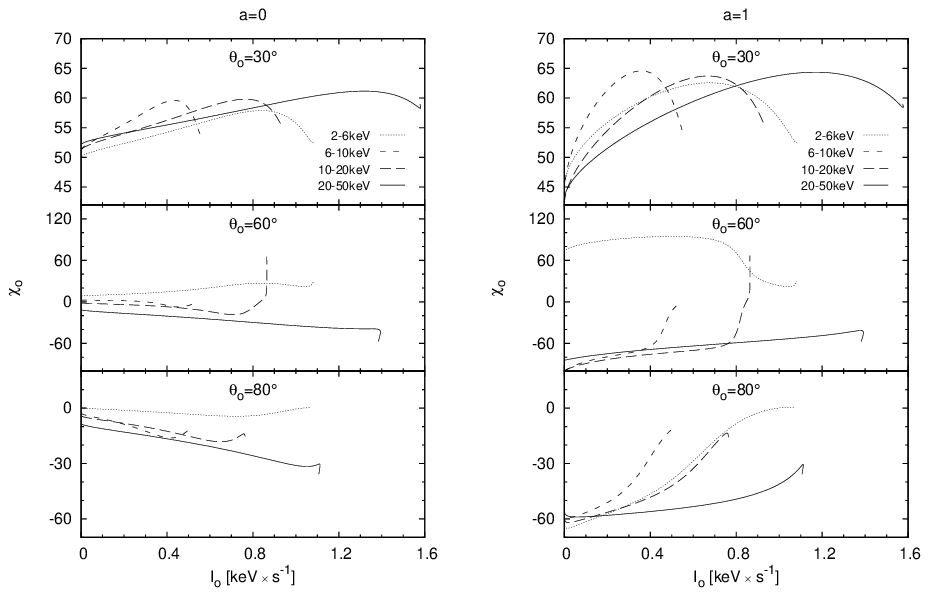}
\end{center}
\caption{Dependence of the polarization angle at infinity on the total
observed intensity. A polarization angle of $0^\circ$ represents the direction
perpendicular to the disk.
The values for the energy ranges 2--6, 6--10, 10--20, and 20--50~keV for three
different observer inclinations, $\theta_{\rm o}=30^\circ$, $60^\circ$, and
$80^\circ$ (top to bottom panels) are shown. The height of the primary source
increases with the total intensity from $0$ to $100\,GM/c^2$.
{\em Left:} The non-rotating Schwarzschild black hole case (a=0).
{\em Right:} The maximally rotating Kerr black hole case (a=1).}
\label{fig:pol_angle_itot}
\end{figure*}

The dependence of the polarization degree and angle at infinity on the
lamp-post height, observer inclination, and total intensity at infinity
are shown in Figures~\ref{fig:pol_degree_hie}--\ref{fig:pol_angle_itot}.
In all these figures both the reflected and primary emission are
considered and the results are computed in several energy bands.

To understand the relativistic results one has to take into account several
aspects different from the Newtonian case:
\begin{enumerate}
\item {\em Local polarization.}
The emission angle, on which the local polarization properties depend,
is not equal to the observer's inclination. Its value depends on the position
on the disk from which the photon is emitted. The emission angle is approaching
the observer inclination only for large radii. Therefore if the local
polarization depended only on the emission angle, it would be the same all over
the disk in the Newtonian case whereas it would change considerably in the
relativistic case. The local incident angle and relative azimuthal angle between
the incident and emitted light rays are different as well.
\item {\em Transfer function.}
The transfer function in the eqs.~(\ref{stokesI})--(\ref{stokesU}) acts as a
weight that amplifies the emission in some areas on the disk while in other
parts it can very efficiently suppress the outcoming radiation. Due to
the relativistic effects this impact is non-axisymmetric. Thus whereas in the
Newtonian case the overall polarization angle may be parallel or perpendicular
due to the symmetry of the system, this does not apply to the relativistic case
where the observed polarization angle may, in principle, attain any value.
\item {\em Energy dependence.}
While in the Newtonian case the energy dependence of the local polarization does
not play any important role it is not so in the relativistic case. Due to energy
shift of the photons the local intensities in the
eqs.~(\ref{stokesI})--(\ref{stokesU}) work as weight similarly as the transfer
function mentioned above. Thus the features such as the unpolarized iron line,
the iron edge and the Compton hump may change the contribution of different
parts of the disk to the polarization in a particular energy band, especially if
those features are shifted in and out of the energy band of interest.
\item {\em The dependence of the relativistic rotation of the polarization angle
on observer inclination.}
To understand this effect one has to realize the role of the critical
point, where the photons are emitted perpendicularly to the disk. The existence
of this point is due to special relativistic aberration. For small inclinations
even very small orbital speed of the matter moving towards the observer is
enough to ensure that light rays emitted perpendicularly to the disk reach the
observer. For very large inclinations the orbital speed must be much higher.
Therefore the critical point is far away from the center for small observer
inclinations whereas it moves close towards the center for very high ones.
In the latter case the contribution of the general relativistic aberration and
light bending becomes important and as a result the location of the critical
point is behind the black hole. For small inclinations the critical point is at
the azimuth where the disk is moving towards the observer.
For the importance of the special relativistic effects (aberration, Doppler
effect and thus special relativistic rotation of the polarization angle) and the
general relativistic effects (significant for small radii: light bending,
gravitational shift, and general relativistic rotation of polarization angle)
see Figure~3 in \cite{dovc08}, Appendix D in \cite{dovc04}, or Figure~1.3 in
\cite{dovc10}. One can see there that for radii below
the critical radius the relativistic rotation of the polarization angle, $\psi$,
spans the whole possible range (i.e. from $-180^\circ$ to $+180^\circ$) whereas
above the critical radius the rotation is restricted to some interval that gets
quite narrow for large radii.
\item {\em Spin of the black hole.}
The disk around the black hole is supposed to terminate at the marginally stable
orbit which means there is a cavity in the disk around the black hole; the lower
the black hole spin, the larger the cavity.
Because the largest influence of the relativistic effects are visible mainly in
close vicinity to the black hole, the relativistic effects are much more
pronounced for highly spinning black holes.
\end{enumerate}

As a consequence, different parts of the disk will emit light with
different local polarization. The direction of this polarization will be further
changed as the photon travels from local frame co-moving with the disk to the
stationary observer and this change will be again dependent on the place of the
emission. The contribution to the overall polarization from each part of the
disk will be different depending on the weight given by the transfer function
and the local intensity (both dependent on the position on the disk). The latter
one is proportional to the incident intensity from the primary source thus the
height of the lamp determines at which radii the disk shines the most. The
incident angle under which photons strike the surface of the disk also depends
on the height of the lamp, so the height modifies the local polarization degree
and angle as well.

Thus while in the Newtonian case it is quite straightforward to see how the
polarization by reflection from the disk behaves it is much more complicated in
the relativistic case. Therefore, in the following, we will restrict ourselves
to the description of the most interesting results and we will discuss only the
main qualitative features.

We begin with a bit surprising and contra-intuitive result ---
the polarization degree is highest when the radiation from the inner
accretion disk has a large contribution to the overall flux.
This is opposite to the de-polarization one gets in the inner disk for the
Comptonized thermal radiation, \citep[see][]{dovc08,schn09}. The cause of the
thermal de-polarization is the relativistic rotation of the polarization angle.
The local polarization is in this case always either parallel or perpendicular
to the disk. In the reflection model the local polarization angle depends on the
local scattering geometry. When integrating the contributions from the disk, the
local polarization angle adds up with the relativistic rotation in such a way
that the overall polarization is much less de-polarized.

This has two
consequences. Firstly,
the highest polarization degrees are achieved for low lamp-post heights when the
inner region of the disk is more illuminated than its more distant parts.
Secondly, the radiation from the extreme Kerr black-hole accretion disk is
more polarized than that in the Schwarzschild case where there is a hole in the
disk below marginally stable orbit. Note, however, that there are
some exceptions to the latter behavior, see the results for some intervals of
the lamp-post heights and for the  observer inclination $30^\circ$ and
$80^\circ$. This can be explained by the fact that the polarization in
the Schwarzschild case is determined only by the disk above marginally
stable orbit, $r>r_{\rm ms}=6\,GM/c^2$, while in the Kerr case the contribution
from the disk below and above this radius have different polarizations,
effectively reducing the overall polarization.

Note, that in Kerr case the dependence of the polarization degree on the
height, Figure~\ref{fig:pol_degree_hie}, roughly follows the dependence of the
ratio of the reflected and primary radiation, $I_{\rm refl}/I_{\rm prim}$,
see middle panel in Figure~\ref{fig:intensity_ref_integrated}. This is not true
for the Schwarzschild black hole where for low lamp-post heights the
polarization is quite large in spite of the fact that the reflected intensity
is much smaller (due to the missing innermost part of the disk). This, again, is
just another consequence of the fact that the
contribution of the inner disk above the marginally stable orbit to the overall
observed polarization is significant.

We should also
mention that similarly to the intensity also the polarization degree and
angle do not change with the spin for large heights ($h\gtrsim 30\,GM/c^2$)
of the primary source (see left panel in Figure~\ref{fig:pol_degree_hie} and
both panels in Figure~\ref{fig:pol_angle_he}).

The polarization degree has a maximum for low observer inclinations, e.g.
in the highest studied energy band, $20-50\,$keV, for the lamp-post height
$h=3\,GM/c^2$ it peaks in the extreme Kerr
case at inclination $25^\circ$ where it reaches almost $15\%$
polarization (see the middle panel in Figure~\ref{fig:pol_degree_hie}).

The polarization degree for zero inclination angle has to be zero because of the
symmetry. As soon as the symmetry is broken by non-zero inclination the total
polarization increases, determined by the region below critical radius which for
low inclinations is far from the center. As soon as the emission above this
radius starts to contribute significantly the overall polarization starts to
decrease with the inclination angle (the larger the inclination the
lower the critical radius). This turn-around is at lower inclinations for larger
heights because higher lamp-post illuminates better farther radii (i.e. radii
above the critical one). There is another turn-around at the inclination when
the critical radius moves close to the center. The contribution
to the polarization from below and above critical radius cancel each other and
the dependence of the polarization degree on the observer inclination reaches
its minimum. For even higher inclinations the polarization again increases, it is
determined mainly by the emission from far above the critical radius with lower
and lower contamination from the regions around and below this radius. For very
high inclinations the reflection is small and thus the overall polarization
decreases again.

The polarization angle at infinity is quite sensitive to the details described
in the points mentioned above and its dependence on height and inclination is
rather complex. Note mainly the different behavior in different energy bands
(Figures~\ref{fig:pol_angle_he}--\ref{fig:pol_angle_ie}). While the polarization
degree behaves in different energy bands very similarly, just being scaled up
with higher energy, the polarization angle dependences may differ substantially.
As discussed above for the polarization degree, the polarization angle for low
inclinations is also determined mainly by the region below critical radius
(top panels in Figure~\ref{fig:pol_angle_he}) while for high inclinations
by the region above this radius (middle and bottom panels in the same figure).
The change in behavior is very well visible by rapid change of polarization
angle with inclination in Figure~\ref{fig:pol_angle_ie}. The transition happens
at the inclination when the polarization above critical radius starts to
dominate for higher inclinations. The transition depends on the lamp-post height
and energy band and it can be either gradual or quite abrupt.
Notice that there is no transition (or a very mild one) in the bottom panel in
Figure~\ref{fig:pol_angle_ie} for the extreme Kerr black hole and for a very low
height of the primary source. This is due to the fact that the inner region
below the critical point still has large impact on the polarization at infinity
even for very high inclinations. In this case also the dip in the dependence of
the polarization degree on inclination is not so deep, see the dotted graphs of
bottom middle panel in Figure~\ref{fig:pol_degree_hie}.

For very high inclinations the
polarization angle at infinity is determined by outer regions of the disk. Here
relativistic effects are small enough so that the polarization at infinity is
given by the local values in the disk. Because the incident and emission angles
are very
large in this area and for this inclination, the polarization angle will be
$0^\circ$ (or $180^\circ$). The values in Figure~\ref{fig:pol_angle_ie} are
smaller, mainly for high heights because the disk in our computations does not
extend to infinity but up to $1000\,GM/c^2$ and therefore the incident angle is
smaller than $90^\circ$ and the local polarization angle decreases.

It is good to understand the behavior of the polarization for different
lamp-post heights and observer inclinations, however, the height
cannot be directly measured and to test the dependence on inclination we would
need polarimetric observations of many AGNs. Therefore in the next sections
we will use the dependence of the polarization on the directly measurable total
intensity that can be observed at infinity. To this purpose we show the
theoretical curves for different energy bands for the polarization degree in the
right panel in Figure~\ref{fig:pol_degree_hie} and for the polarization angle in
Figure~\ref{fig:pol_angle_itot}. In these figures the height of the primary
source increases with the total intensity.

\section{Observational perspectives}
\label{sec:observations}
\subsection{Monte Carlo simulations of the detector response}

The complex dependence of the polarization on the geometry of the system
gives important insight in the emission process at work in accreting
black holes. Many missions with the capability of measuring polarization
are expected or are discussed for launch in the next few years and
therefore it is important to investigate if the sensitivity is
sufficient to perform polarimetry also of relatively faint sources like
AGN. For this purpose, we developed
a dedicated Monte Carlo software which is described below.

Instruments able to measure polarization exploit the modulation which
appears in the case of polarized radiation in the azimuthal response.
Experimentally, this response, or modulation curve, is the histogram of
the initial direction of photoelectron for photoelectric polarimeters
(\citeauthor{Bellazzini2010}, \citeyear{Bellazzini2010}) or that of the
scattering direction for Compton devices (\citeauthor{McConnell2010},
\citeyear{McConnell2010}). If the impinging photons are polarized, the
photoelectrons (or the scattered photons in the case of Compton
instruments) are not isotropically distributed, but their azimuthal
distribution on the plane orthogonal to the direction of incidence shows
a cosine square modulation. The peak (or the minimum) corresponds to the
direction of polarization, while the amplitude of the modulation is
linearly proportional to the degree of polarization. The maximum
modulation achieved for 100\% polarized radiation is called modulation
factor of the instrument $\mu$ and it is in general a function of the
energy.

In the following we will assume instruments based on a photoelectric
polarimeter, the Gas Pixel Detector (GPD; \citeauthor{Bellazzini2010b}
\citeyear{Bellazzini2010b}). The
systematic effects of GPD have been reported to be below 0.3\%
\citep{Bellazzini2010}, and therefore they can be neglected if the
polarization degree is above $\sim$1\%. In this hypothesis, the sensitivity of
a polarimeter is only a matter of how many photons are collected. The
number of photoelectrons emitted in each azimuthal angular bin is
Poisson distributed and its fluctuations ultimately limit the capability
to measure the amplitude of the modulation curve and, consequently, the
polarization \citep{Weisskopf2010}.

The basic idea behind our Monte Carlo software is to derive the degree
and the angle of polarization by generating test modulation curves, that
is trial histograms representing the azimuthal distribution of
photoelectrons. The entries in each histogram, one for each spectral
interval of interest, is the number of collected photons in that energy
band. The number of counts $N_i$ in each angular bin $i$ is distributed
as a cosine square function, whose phase and amplitude are the angle of
polarization and the product of the modulation factor and the degree of
polarization averaged in the energy band, respectively. A Poisson noise
$\sqrt{N_i}$ is added in each angular bin. The modulation curve is
fitted with a function $\mathcal{M}(\phi)=A+B\,\cos(\phi-\phi_0)$, where
$A$ and $B$ are constants. The fit returns the ``measured'' angle of
polarization $\Phi_j=\phi_0$ and its error $\sigma_{\Phi j}$, while the
polarization degree $\mathcal{P}_j$ (alongside with its error) is
derived from the usual formula
\begin{equation}
\mathcal{P}_j =
\frac{1}{\mu}\,\frac{\mathcal{M}_{max}-\mathcal{M}_{min}}{\mathcal{M}_{max}+
\mathcal{M}_{min}} = \frac{1}{\mu}\,\frac{B}{B+2\,A}.
\label{eq:Pol}
\end{equation}
The fit is repeated $J$ times, e.g. 1000, adding a different Poisson noise
contribution to the histogram for each trial. Eventually the
expected (``measured'') degree and angle of polarization (and their
errors) are derived by averaging the values obtained by the $j$-th trial,
\begin{equation}
\mathcal{P} = \overline{\mathcal{P}}\equiv\frac{1}{J}\sum_j \mathcal{P}_j,\ \
%\hspace{1cm}
\sigma_\mathcal{P} = \overline{\sigma}_\mathcal{P},\ \
%\hspace{1cm}
\Phi = \overline{\Phi}_j,\ \
%\hspace{1cm}
\sigma_\Phi = \overline{\sigma}_{\Phi j}.
\label{eq:PolMean}
\end{equation}

This method was used by \cite{dovc08} to derive the sensitivity of next
X-ray polarimetry missions to the rotation of the plane of polarization
in the emission from accreting Galactic black holes. In that case
polarization was to be measured in quite narrow energy bands, about
1~keV, and therefore it was possible to define the average quantities
required to generate the test modulation curves, e.g. the modulation
factor, the angle and the degree of polarization. Instead here we need
to study the sensitivity on a wider energy interval and therefore we
present a different approach, initially proposed by Costa (private
communication), which has also the advantage of taking into account the finite
energy resolution of the instrument.

The simulation starts from the knowledge of the modulation factor, the
spectrum of the source, the expected degree and the expected angle of
polarization. They are calculated in $K$ narrow spectral bins, typically
0.1 keV which is well below the energy resolution of the Gas Pixel
Detector (about 1~keV at 6~keV). The values of the modulation factor, of
the spectrum, of the degree and of the angle of polarization in the
$k$-th spectral bin $E_k$ will be named $\mu_k$, $F_k$, $P_k$,
$\varphi_k$, respectively. For each $k$-th bin in energy, we build a
histogram with a cosine square modulation proportional to $P_k\,\mu_k$
and a phase $\varphi_k$. The histograms have $I$ angular bins between 0
and 2$\pi$, e.g. 72 bins corresponding to intervals of 5~degrees, and
the total number of elements is proportional to $F_k$. The value of the
$k$-th element in the $i$-th angular bin $H_{ki}$ is basically the
fraction of the flux in the spectral bin $E_k$ which causes a
photoelectron to be reconstructed in the $i$-th angular direction
(neglecting statistical fluctuations). So far this is not very different
from the approach formerly presented, with the only difference that we
define histograms in very narrow energy bands.

\begin{figure*}[tbh!]
\begin{center}
\includegraphics[width=\textwidth]{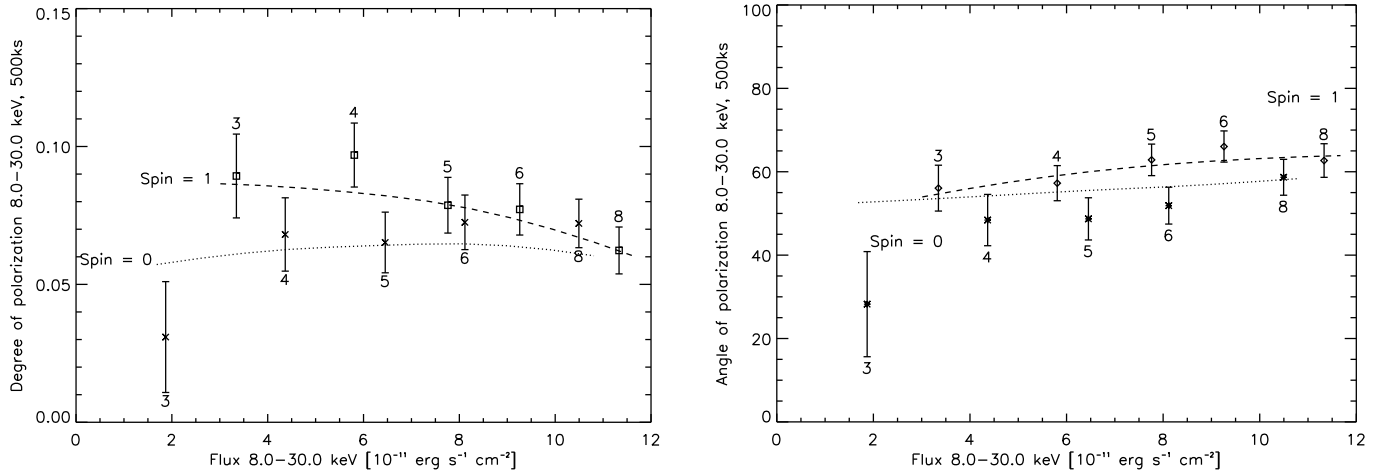}
\end{center}
\caption{Correlation of the degree (left) and of the angle of
polarization (right) with the flux for MCG-6-30-15. Each point
corresponds to a different height of the source, specified
within the figure (in units of $GM/c^2$). It is assumed that each
state of the source is observed for 500~ks with the Medium Energy
Polarimeter on-board NHXM in the 8--30~keV energy interval. The dashed
and the dotted lines are the expected dependency for spin 1 and spin 0,
respectively.}
\label{fig:MCG}
\end{figure*}

The finite spectral resolution of the instrument makes the measurement
of the energy of the photon to be smeared around an average value. This
causes a photon at energy $E$ to be reconstructed as a photon at
different energy $\hat{E}$, and the difference between $E$ and $\hat{E}$
depends on the energy resolution. The uncertainty on the energy
measurement needs to be convolved with the response of the instrument,
in particular with the dependence of the modulation factor on energy.
As a matter of fact, completely polarized photons at energy $E$ cause a
modulation of amplitude $\mu(E)$ which can be quite different from
$\mu(\hat{E})$. To model properly the amplitude of the modulation, one
should take into account that the instrument responds to polarization
differently with the energy and that the energy is measured with a
finite resolution. In our Monte Carlo simulations we include this degeneracy. We
assume, and this is appropriate in most cases, that the effect of the
finite energy resolution is limited to smear the energy of the event and
that it does not act on the direction of emission of the photoelectron,
namely on the polarization. Therefore we apply the blurring due to the
finite spectral resolution to the spectra resolved in angular bins.
Practically, this is done by taking the $K$ values $H_{k\overline{i}}$ for any
angular bin $i=\overline{i}$ and for $k\in[0,K-1]$, where we recall
that $k$ is the index running over the spectral bins. These $I$ spectra
are introduced in XSPEC as additional models and the response matrix of
the instrument is used to apply a Gaussian smearing to them with the
XSPEC tool \texttt{fakeit}. The results are $I$ \texttt{pha} files which
contain the number of counts $N_{ki}$ collected in the angular bin $i$
for each spectral bin $E_k$ in the detector energy space.

The arrays $N_{k'i}$ for any fixed value of $k'$ and any value of $i$ represent
a sort of modulation curves defined in very narrow energy bands $E_{k'}$ of
0.1~keV that we name ``pseudo modulation curve'' (because the Poisson
noise has not been included yet). Any practical measurement will be
performed in much larger energy intervals, of the order of a few keV.
Therefore we build the pseudo modulation curves in any energy band of
interest $\Delta E_l$ by summing those histograms $N_{ki}$ which are
referred to spectral bins $E_k$ contained in the interval $\Delta E_l$.
This results in $L$ pseudo modulation curves, as many as the number of
intervals, which we name $M_l$, where we drop the subscript $i$ (we
implicitly assume the angular dependence being indicated by $i$). The values
of $M_l$ take into account the effect of the finite energy resolution on
the amplitude of the modulation and that the degree and the angle of
polarization can change with energy, that is these quantities are
assumed constant only on small intervals of 0.1~keV.

After that the pseudo modulation curves are generated, the Monte Carlo
scheme proceeds as discussed above. In each $\Delta E_l$ energy band, the
software produces $J$ trial modulation curves, which are calculated from
the same pseudo modulation curves $M_l$ but adding for each trial a
different Poisson noise. Each trial modulation curve is fitted with the
function $\mathcal{M}(\phi)=A+B\,\cos(\phi-\phi_0)$, and eventually the
``measured'' degree and angle of polarization are derived by averaging
the values obtained by each trial; see equation (\ref{eq:PolMean}). The
value of the modulation factor used in each energy band to derive the
polarization degree from the amplitude of the modulation (see
equation~\ref{eq:Pol}) is obtained by weighting the dependency of $\mu$
with energy on the spectrum of the source.

\subsection{Results}

The Monte Carlo scheme was used to
investigate if polarimeters on-board next missions could detect
distinctive signatures of the lamp-post model. A key characteristic of
such a model is that the height of the illuminating source must change
to explain the temporal variation of the source. This results also in a
change of the observed flux and polarization, as reported in right panel in
Figure~\ref{fig:pol_degree_hie} and \ref{fig:pol_angle_itot}.
Therefore our primary objective is to
discuss if next missions will be able to find an evidence of the
positive correlation between the degree of polarization and the flux.

\subsubsection{Predictions for AGN case: MCG-6-30-15}
First we model the case of Seyfert 1 galaxy MCG-6-30-15.
From the simulations presented above we know how the flux and the
polarization correlate, but we still need to associate the primary
source height
to a total flux measured in a certain state. Since we can not
derive this normalization from the available information, we assume that
the lowest flux state measured by \cite{Lee2000} corresponds to
$h=3\,GM/c^2$. This choice, though arbitrary, is
supported by conclusions of other authors, who suggest
that the illuminating
source is quite close to the central black hole. We assume that the
observation inclination is 30$^\circ$ and that the black hole in
MCG-6-30-15 is maximally rotating.

An interesting result expected on the basis of the lamp-post model is
that the polarization usually increases with energy (see
Figure~\ref{fig:polar_ener}). Therefore in the following we will take as
an example the New Hard X-ray Mission (NHXM hereafter), which is
dedicated to perform imaging, spectroscopy, and polarimetry in a wide
energy interval extending from soft to hard X-rays
(\citeauthor{Tagliaferri2010}, \citeyear{Tagliaferri2010}). NHXM will
allow for polarimetry above 10~keV where the reflected radiation start
to dominate the emission. In the focus of one of the four multilayer telescopes
there will be alternatively placed two photoelectric polarimeters
based on the Gas Pixel Detector, which are sensitive in the intervals
2--10 and 6--35 keV, and possibly a third Compton polarimeter which would
extend the response up to 80~keV \citep{Soffitta2010}.
Here we focus on the Medium Energy Polarimeter
(MEP), the polarimeter that will be sensitive between 6 and 35~keV.

\begin{figure*}[tbh!]
\begin{center}
\includegraphics[width=\textwidth]{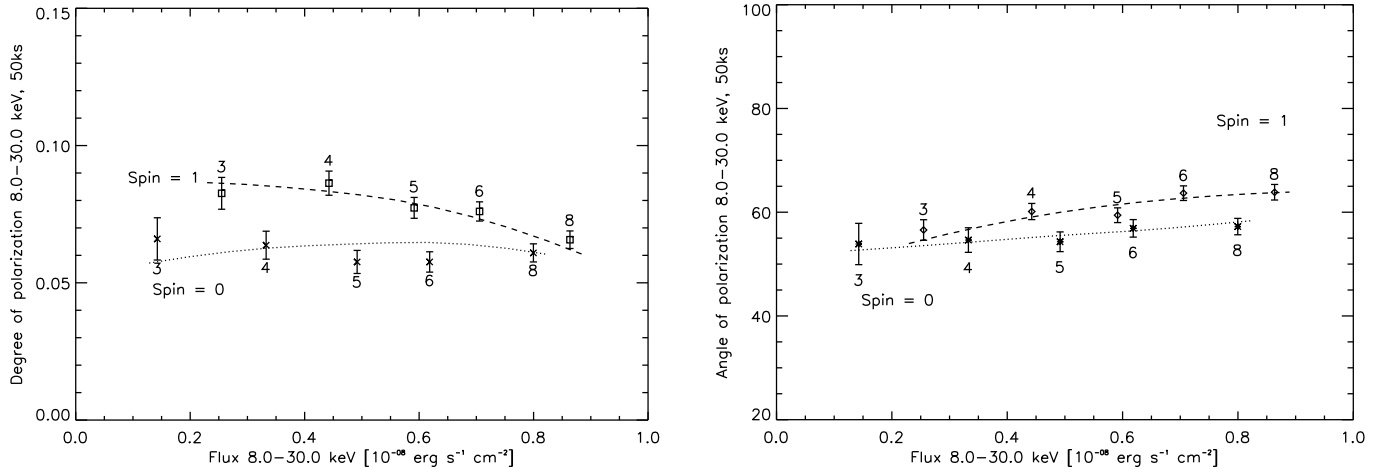}
\end{center}
\caption{The same as Figure~\ref{fig:MCG} but for the Galactic black hole
system XTE~J1650-500. It is assumed that each state of the source is
observed for 50~ks with the MEP on-board NHXM in the 8--30~keV energy
interval.}
\label{fig:J1650}
\end{figure*}

We show  the correlations of the degree and of the angle of polarization with
the flux in Figure~\ref{fig:MCG}. Both the flux and the polarization are
integrated in the energy
range 8--30~keV to avoid the iron line which is expected to be
unpolarized. Each point represents an observation of 500~ks and
corresponds to a different state of the source, that is a different
height as indicated in the figure. Although the angle of polarization is
almost constant with the flux, a certain evolution of the degree of
polarization should be detectable. Moreover, we report for comparison
the evolution of the polarization expected in the case of a
Schwarzschild black hole to show that such evolution is somehow
dependent on the spin of the central black hole.

\subsubsection{Predictions for Galactic black hole systems: XTE J1650-500}
Features which can be regarded as due to reflection, such as
relativistic iron lines, are observed in a number of Galactic black hole
systems; see \cite{Mill07} for a review. There have been attempts to apply
the light bending scenario also to these sources in analogy with the AGN
case and therefore we discuss also the possibility to observe the
correlation between the polarization and the flux in the case of
Galactic objects.

We take as an example the transient black hole system
XTE~J1650-500 and associate the maximum flux measured by \cite{Rossi05}
to the a source height $h=8\,GM/c^2$, changing the source height
between this value and $h=3\,GM/c^2$. We assume that the source
spin is 1 and the observer inclination is 30$^\circ$.
We suppose to observe each state of XTE~J1650-500 for 50~ks with the MEP
on-board NHXM. The result is reported in Figure~\ref{fig:J1650}. In this
case a significant detection could be reached with quite short
observations.

\section{Discussion and conclusions}
\label{sec:conclusions}
In this paper we have discussed the observational properties of the intensity
of the emerging primary radiation component, the radiation component
incident on the accretion disk, and the reflected
radiation in the lamp-post geometry (the light bending model). We have extended
our previous work on the subject and showed new results for
polarization computations, including the unpolarized iron line complex
in $2$--$10\,$keV energy range and the Compton hump at higher
energy band (above $15$~keV). We used our theoretical computations to
model possible future observations by next-generation X-ray satellite
missions equipped with an X-ray polarimeter on board.

For low inclinations we find that the polarization of the reflected
radiation is diminished in the broad iron line energy region. For
high inclinations, however, the situation is more complicated due to
the shift of the iron edge; this effect enhances the observed
polarization and it may balance the expected decrease. In principle the effect
of lower polarization of the relativistic broad line region could be used
to discriminate between the relativistically broadened iron line and
the partial covering scenario, which are both proposed to explain the
excess of flux in the
spectra of black-hole accretion disks between 2--10~keV
\cite[see e.g.][]{mill09}. Note, however, that we did not try to estimate the
behavior of the polarization in the partial covering scenario in this paper.

The polarization degree in a lamp-post geometry is higher at the
highest studied energy band, 20--50~keV. This is a natural result
coming from the fact that the primary source has a power-law spectrum
with a negative index. Thus the best results for polarimetry of reflection
spectra should be achieved in the Compton hump energy region. We also
get higher polarization degree for large values of spin of the black hole, small
height of the primary source, and low inclination of the observer.

As an example of the polarimetric sensitivity of next missions, we could
assume the performance of the Gravity and Extreme Magnetism SMEX (GEMS;
\citeauthor{Jahoda2010}, \citeyear{Jahoda2010};\citeauthor{Black2010},
\citeyear{Black2010}), which will be launched by 2014.
However, the energy range of the polarimeters on-board GEMS will be limited
between 2 and 10~keV, with the peak of the sensitivity being around $3-4$ keV.
Thus it will observe photons in a region where the iron (unpolarized)
fluorescent line is prominent and the direct radiation from the corona largely
dominates the reflected  component. The reflected radiation is the only one
polarized in our model. Consequently the expected degree of polarization is
quite low, between 1
and 2\% depending on the height of the illuminating source. Thus the signal
could be more easily affected by even a small amount of polarization of some
other origin, e.g. by Comptonized thermal emission in case of X-ray Black hole
binaries (\citeauthor{dovc08}, \citeyear{dovc08};
\citeauthor{li09}, \citeyear{li09}; \citeauthor{schn09}, \citeyear{schn09}).
Therefore we preferred to simulate observations of the polarization in higher
energy band.

In our model there were several assumptions and approximations that could affect
the resulting polarization. First of all the primary radiation was unpolarized.
It is clear that the effect this will have on our results will be most prominent
in low energy bands, for higher lamp-post heights, low inclinations or small
black-hole spins where primary emission is dominant (see
Figure~\ref{fig:intensity_ref_integrated}). Thus the polarization at low
energies could be enhanced but determined by the primary source while the
polarization at higher energies could be reduced.

In our computations we used the single scattering approximation for local
polarization properties. The accuracy of this approximation depends on
the energy. One can have an idea of the effect confronting Figure~1 (lower
panel) and Figure~3 in \cite{Matt93}. It follows from these figures that for
energy below $10\,$keV results should not change much, while they could be
different by about 25\% in the $10-30\,$keV and by almost 40\% in the
$30-50\,$keV energy range.

We assumed a smooth equatorial disk. In case of a ``rough'' disk, with random
ripples of the length much smaller than the length at which relativistic effects
change considerably, the local polarization properties should be averaged over
incident and emission angles. Because we average over these values anyway when
integrating over the disk but including the relativistic effects it is hard to
predict in a unique way how this
will change the overall polarization at infinity. In case the disk is smooth but
not strictly equatorial (warped disk but still close to the equatorial plane),
the local geometry of scattering in the disk will be changed and thus the
polarization at infinity will be changed as well, both dependent on the geometry
of warping. In case of largely warped disks or non-aligned
disks we expect different results mainly because the disk would be more
illuminated at places closer to and facing the primary source, and because of
the different local geometry of scattering and different values of the
relativistic effects.

Our work is complementary to \citet{schn10} who assume a different
geometry and structure of the corona. A direct comparison of their paper
with our results is not easy because of different approximations adopted in
these two works and the notorious sensitivity of polarization to details
of the model. Nevertheless, the general set-up is similar in both
scenarios and the level of expected polarization is comparable.

The behavior of the polarization degree and polarization angle in the
light bending model is quite complex. Because the result depends on the
interplay of several parameters, the polarization properties may be
degenerate with respect to different parameter values (e.g.\ for larger
heights we cannot distinguish a difference between a rotating and a
non-rotating black hole). Nevertheless, when combined with spectral and
timing observations, polarimetry is yet another important channel that
can help us to uncover the physical parameters of the black-hole
accretion disk systems, such as the black hole spin, the system
inclination, and the height of the illuminating source.

\medskip
\section*{Acknowledgments}
The authors would like to thank anonymous referee for her/his comments that have
lead to substantial improvement of the manuscript.
We thank E.~Costa, P.~Soffitta, R.~Bellazzini, and A.~Brez for
making available the sensitivity model of the Gas Pixel Detector for NHXM
mission, and S.~Bianchi for useful discussions. MD and VK acknowledge the Czech
Science Foundation (ref.\ 205/07/0052) and the ESA PECS project (ref.\
98040). The Astronomical Institute has been supported by the project
LC06014 of the Centre for Theoretical Astrophysics in the Czech
Republic. GM and FM acknowledge financial support from Agenzia Spaziale
Italiana (ASI I/088/06/0). MD and RG are grateful for the support by the
European Researchers Exchange program organized by the AS CR and the French
CNRS.

%\appendix
%\section{...}
%\label{appa}
%\clearpage

\end{document}